\documentclass[submission,copyright,creativecommons]{eptcs}
\providecommand{\event}{ICLP 2020} 
\usepackage{underscore}           

\usepackage{graphicx}
\usepackage{mathptmx}
\usepackage{tikz-qtree}
\usepackage{pgfplots}
\usepackage{filecontents}
\usepackage{csvsimple}
\usepackage{amsfonts,amsmath}
\usepackage{url}
\usepackage{verbatim}
\usepackage{color}

\definecolor{lgray}{gray}{0.95}
\definecolor{lblue}{rgb}{0.90,0.90,1.00}
\definecolor{lyellow}{rgb}{1.00,1.00,0.70}

\usepackage{listings}
\newtheorem{ex}{Example}

\newenvironment{codex}{\small\verbatim}{\endverbatim\normalsize}

\lstnewenvironment{code}
    {\lstset{}%
      \csname lst@SetFirstLabel\endcsname}
    {\csname lst@SaveFirstLabel\endcsname}
\newcommand{\BI}[0]{\begin{itemize}}
\newcommand{\EI}[0]{\end{itemize}}
\newcommand{\I}[0]{\item}
\newcommand{\BE}[0]{\begin{enumerate}}
\newcommand{\EE}[0]{\end{enumerate}}

\newcommand{\BX}[0]{\begin{ex}}
\newcommand{\EX}[0]{\end{ex}}
\newcommand{\BV}[0]{\begin{verbatim}}
\newcommand{\EV}[0]{\end{verbatim}}

\newcommand{\BF}[0]{\begin{filecontents*}{data.csv}}

\newcommand{\BQ}[0]{\color{blue}\begin{quote}}
\newcommand{\EQ}[0]{\end{quote}\color{black}}

\def \bscale1 {0.25}
\def \bscale {0.25}

\usepackage{amssymb}
\usepackage{hyperref}
\usepackage{booktabs}

\begin{document}

\title{Deriving Theorems in Implicational Linear Logic, Declaratively} 

\author{Paul Tarau
\institute{University of North Texas\\ 
Texas, USA
}
\email{paul.tarau@unt.edu}
\and
Valeria de Paiva  
\institute{Topos Institute\\
California, USA}
\email{\quad valeria.depaiva@gmail.com}
}

\def \authorrunning{Paul Tarau and Valeria de Paiva}
\def\titlerunning{Deriving Theorems in Implicational Linear Logic, Declaratively}

\maketitle

\begin{abstract}
The problem we want to solve is how to generate all theorems of a given size in the implicational fragment of propositional intuitionistic linear logic.
We start by filtering for linearity the proof terms associated by our Prolog-based theorem prover for Implicational Intuitionistic Logic. This works, but using for each formula a PSPACE-complete algorithm limits it to very small formulas. We take a few walks back and forth over the bridge between proof terms and theorems, provided by the Curry-Howard isomorphism, and derive step-by-step an efficient algorithm requiring a low polynomial effort per generated theorem. The resulting Prolog program runs in $O(N)$ space for terms of size $N$ and generates in a few hours 7,566,084,686 theorems in the implicational fragment of Linear Intuitionistic Logic together with their  proof terms in normal form. As applications, we generate datasets for correctness and scalability testing of  linear logic theorem provers and training data for neural networks working on theorem proving challenges. The results in the paper, organized as a literate Prolog program,  are fully replicable.

\end{abstract}

\section{Introduction}

{\em Linear Logic} \cite{girard1987} as a resource-control mechanism constrains the use of formulas available as premises in a proof. In its full generality, a larger number of operators ensures on-demand (re)use of these resources, in a controlled way (e.g., with exponentials like ``{\bf !}''). While in full propositional form linear logic is already Turing complete, its {\em implicational fragment} is decidable and finding low polynomial algorithms for proving its theorems is especially interesting when large  datasets of theorems need to be generated. Such datasets, combining tautologies and their proof terms can be useful for testing correctness and scalability of linear logic theorem provers (not necessarily restricted to the implicational fragment) and more importantly, for training   deep learning networks focusing on {\em neuro-symbolic} computations, e.g., \cite{neuralLMs,deepProblog,difProving}, an emerging research trend, motivated in part by the need for {\em explainable AI} in medical, legal or other industrial AI applications.

Of particular interest in the correspondence between computations and proofs is the Curry-Howard isomorphism \cite{howard:formulaeastypes:hbc:80,wadler15}.
In its simplest form, it connects the implicational fragment of propositional intuitionistic logic with types in the {\em simply typed lambda calculus}. A low polynomial type inference algorithm associates a type (when it exists) to a lambda term. Harder (PSPACE-complete, see \cite{statman79}) algorithms associate inhabitants to a given type expression  with the resulting lambda term (typically in normal form) serving as a witness for the existence of a proof for the corresponding tautology in  implicational propositional intuitionistic  logic.
In particular, when restricting linear logic to its implicational fragment (syntactically, just binary trees with the ``lollipop'' operator ``\verb~-o~'' and variables as leaves), it becomes interesting to find out how formulas  relate to proof terms, seen as linear lambda terms (constrained to have exactly one variable associated to each lambda binder).
Also, this is important because such formulas correspond to linear types, which can significantly optimize memory management by allowing reuse of single-threaded data structures as it has been implemented in Linear Haskell  \cite{linhask}.
 
 This singles out the usefulness of efficiently generating a dataset of linear types/linear logic tautologies, the focus of this paper, with at least three applications in mind:
 
 \BI
 \I correctness and scalability tests for linear logic theorem provers, complementing the ones described in \cite{olarte2018}
 \I a formula/proof term dataset for training neuro-symbolic systems with a likely to be learnable, PTIME-decidable set of problems
 \I a correctness and scalability test for systems implementing linear types (e.g., Linear Haskell)
\EI

We will proceed incrementally, with a step-by-step derivation process, starting with adapting an intuitionistic theorem prover to work as a prover for the implicational fragment of linear intuitionistic logic. 
From this solution, seen as an executable specification (correct but slow) we derive, after crossing the Curry-Howard ``bridge'', progressively more constrained lambda term generators, ending with one that not only generates efficiently closed linear lambda terms in normal form but it also infers their types, corresponding to theorems in the language of implicational linear intuitionistic logic. Moreover, we engineer the generation mechanism such that the lambda terms and their principal types have exactly the same size. Thus, without help of a theorem prover, we will uniformly generate all linear implicational tautologies of a given size
\footnote{A dataset containing the theorems generated
and their proof-terms is available at
\url{http://www.cse.unt.edu/~tarau/datasets/lltaut/} .}.
As a result, our  Prolog code defines constructively a size-preserving bijection between these two sets, on the opposite side of the Curry-Howard bridge. As a final step, we re-engineer this bijection to work in reverse mode, as a theorem prover, that given an implicational formula, returns its proof term, if it exists.

The rest of the paper is organized as follows.
Section \ref{fgen} introduces formula generators for (linear) implicational formulas of a given size.
Section \ref{prover} describes the adaptation of an intuitionistic theorem prover to formulas of implicational propositional linear logic.
Section \ref{bridge} moves our effort to the other side of the Curry-Howard isomorphism,
resulting in generation of linear lambda terms in normal form that are bijectively connected to
their principal types corresponding to theorems in implicational linear logic.
Section \ref{disc} discusses  our results in the wider context of linear logic research. 
Section \ref{rel} overviews related work and section \ref{concl}  concludes the paper. {\em As the paper is actually a literate Prolog program, it's code is also made available as a separate file\footnote{\url{https://raw.githubusercontent.com/ptarau/TypesAndProofs/master/tlin.pro}},
in compliance with our commitment to fully replicable research results}.


\section{The Formula Generators}\label{fgen}
We will first develop formula generators to cover all implicational formulas of a given size, measured as the number of internal nodes. With the ``lollipop'' operator ``\verb~-o~'' labeling internal nodes and natural numbers starting with {\tt 0} as variables labeling the leaves, one such formula tree to be generated for N=4, is the following:\\
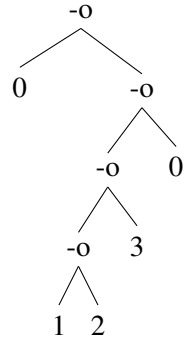
\begin{center}
\Tree [.-o [.0 ] [.-o [.-o [.-o [.1 ] [.2 ]  ] [.3 ]  ] [.0 ]  ]  ]
\end{center}

\subsection{Generating Formula Trees}
First, we generate all binary trees of size {\tt N} with internal implication nodes {``\verb~-o/2~''},  while collecting their {\tt N+1 } distinct logic variable leaves to a list. 

We define ``\verb~-o~'' as an operator and we 
use {\tt pred/2} to consume one unit of 
size on each internal node.

\begin{code}
:-op(900,xfy,( '-o' )).

gen_tree(N,Tree,Leaves):-gen_tree(Tree,N,0,Leaves,[]).

gen_tree(V,N,N,[V|Vs],Vs).
gen_tree((A '-o' B),SN1,N3,Vs1,Vs3):-pred(SN1,N1),
  gen_tree(A,N1,N2,Vs1,Vs2),
  gen_tree(B,N2,N3,Vs2,Vs3).
     
pred(SN,N):-succ(N,SN).
\end{code}

The counts of generated trees match entry {\bf A000108}
in \cite{intseq}, representing the Catalan numbers \cite{StanleyEC}, 
binary trees with $N$ internal nodes.


\subsection{Generating the variable labels}

The next step  toward generating the set of all type formulas
is observing that logic variables define equivalence
classes that  correspond to {\em partitions
of the set of variables}, simply by
selectively unifying them.

The predicate {\tt mpart\_of/2}
takes a list of distinct logic variables
and generates partitions-as-equivalence-relations
by unifying them ``nondeterministically''.
It also collects the unique variables defining
the equivalence classes, as a list given by its second argument.
\begin{code}
mpart_of([],[]).
mpart_of([U|Xs],[U|Us]):-mcomplement_of(U,Xs,Rs),mpart_of(Rs,Us).
\end{code}

To implement a set-partition generator,
we  split a set repeatedly in subset+complement
pairs with help from the predicate {\tt mcomplement\_of/2}.
\begin{code}
mcomplement_of(_,[],[]).
mcomplement_of(U,[X|Xs],NewZs):-mcomplement_of(U,Xs,Zs),
  mplace_element(U,X,Zs,NewZs).

mplace_element(U,U,Zs,Zs).
mplace_element(_,X,Zs,[X|Zs]).
\end{code}
To generate all set partitions from a list of distinct variables
of a given size, we build a list
of fresh variables with 
Prolog's {built-in predicate {\tt length/2}
and constrain {\tt mpart\_of/2} to use them as the set to be partitioned.
\begin{code}
partitions(N,Ps):-length(Ps,N),mpart_of(Ps,_).
\end{code}
The counts of the resulting set-partitions (Bell numbers) ~
{\tt 1, 1, 2, 5, 15, 52, 203,...}
correspond to the entry {\bf A000110} 
in \cite{intseq}.

\BX
Set partitions of size 3 expressed as variable equalities.
\begin{codex}
?- partitions(3,P).
P = [A, A, A]; P = [A, B, A]; P = [A, A, B]; P = [A, B, B]; P = [A, B, C].
\end{codex}
\EX

We next bind leaf variables of formula trees to our set partitions and encode distinct variables as consecutive natural numbers starting at {\tt 0}.
\begin{code}
natpartitions(Vs):-mpart_of(Vs,Ns),length(Ns,SL),succ(L,SL),numlist(0,L,Ns).
   
gen_formula(N,T):-gen_tree(N,T,Vs), natpartitions(Vs).
\end{code}
This sequence corresponds to entry {\bf A289679} in \cite{intseq},
with the first terms being {\tt 1, 1, 2, 10, 75, 728, 8526, 115764, 1776060},
computed as {\tt 	a(N) = Catalan(N)*Bell(N+1)}.
\BX
Some formulas of size 2.
\begin{codex}
?- gen_formula(2,T).
T =  (0 -o 0 -o 0) ; T =  (0 -o 1 -o 0) ;
...
T =  ((0 -o 1) -o 1) ; T =  ((0 -o 1)-o 2) .
\end{codex}
\EX

\section{Adapting a Prover  for Implicational Linear Logic }\label{prover}
We will derive a prover for the implicational fragment of Propositional Intuitionistic Linear  Logic
by adding linearity constraints to the intuitionistic prover described in \cite{padl19},
(also in Appendix A).

\subsection{Ensuring the Proof Terms are Linear}
We will constrain intuitionistic proofs to produce linear lambda terms as proof terms.
Lambda terms are represented using {\tt a/2} for application nodes and logic variables
for binders in {\tt l/2} nodes and the variables they bind.
\begin{code}
is_linear(X) :- \+ \+ is_linear1(X).

is_linear1(V):-var(V),!,V='$bound'.
is_linear1(l(X,E)):-is_linear1(E),nonvar(X).
is_linear1(a(A,B)):-is_linear1(A),is_linear1(B).
\end{code}
The predicate {\tt is\_linear/1} tests that each lambda binder corresponds
to exactly one variable. Double negation is used to undo marking each variable
with the atom ``\verb~$bound~''.

A linear logic prover is now derived from the intuitionistic prover {\tt prove\_ipc}
(see {\bf Appendix})
by filtering {\em proof terms} with {\tt is\_linear}. The predicate {\tt gen\_taut} combines
the implicational formula generator with the linear prover to
obtain  implicational linear logic tautologies of size {\tt N}.
\begin{code}
prove_lin(T,ProofTerm):-prove_ipc(T,ProofTerm),is_linear(ProofTerm).

gen_taut(N,T,ProofTerm):-gen_formula(N,T),prove_lin(T,ProofTerm).
\end{code}
\BX
Formulas of size 3 depicted as trees, together with their proof terms

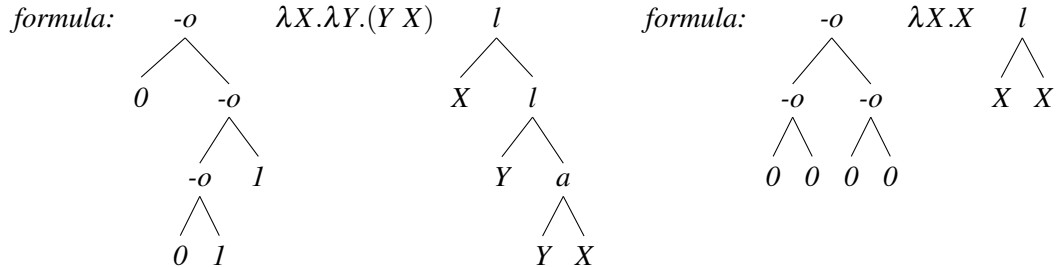
\begin{center}
formula: \Tree [.-o [.0 ] [.-o [.-o [.0 ] [.1 ]  ] [.1 ]  ]  ]
$ \lambda X. \lambda Y.(Y ~X )$
\Tree [.l [.X ] [.l [.Y ] [.a [.Y ] [.X ]  ]  ]  ]
~~~~~
formula: \Tree [.-o [.-o [.0 ] [.0 ]  ] [.-o [.0 ] [.0 ]  ]  ]
$ \lambda X.X $
\Tree [.l [.X ] [.X ]  ]
\end{center}
\EX
This is working but it is too slow, it takes 2203 seconds 
to generate the counts {\tt 0, 1, 0, 4, 0, 27, 0, 315, 0, 5565}.
That's expected, not only because  intuitionistic propositional logic
proofs are PSPACE-complete even for the implicational
fragment, but also because we are filtering through the super-exponential
number of formulas counted by {\bf A289679} in \cite{intseq}.

Besides performance issues, we are facing here three hurdles:
\BI
\I proof terms are not necessarily in normal form
\I multiple proof terms can result in the same provable formula
\I sizes of formulas do  not correlate in a simple way to the sizes of their proof terms
\EI
{\em On the other hand, we know that type inference on lambda terms
resulting in provable formulas can make the process much
faster.} This brings us to our next step.

\section{Crossing the Curry-Howard Bridge: from Lambda Terms to Provable Formulas}\label{bridge}

It's time to  look into the linear lambda terms
corresponding to the formulas we want to generate.

While generators for linear lambda terms do exist (e.g., \cite{lescanne18swiss,padl18} we are starting here with a clean design that propagates {\em size constraints} by keeping separate counts for lambda nodes and application nodes and then enforces linearity efficiently.
This constraint reduces significantly the candidate trees to be decorated with lambda binders
and variables  as it is now like working with size $N$ rather than size $2N+1$ in a super-exponentially growing set, while reducing the possible
leaf labelings to much fewer than all possible combinations of variable names.
Adding  linearity constraints will further reduce the combinatorial explosion by ensuring that each
lambda binder connects to a unique leaf variable.

\subsection{A Generator for Linear Skeleton Motzkin Trees}

First we use the fact that there are as many lambda binders as variables, given the one-to-one mapping required for the (completely) linear lambda calculus. Thus we will give $N$ units to application nodes, corresponding to the $N+1$ variables in leaf position and $N+1$ units to lambda nodes, resulting in a total of $N+N+1=2N+1$ internal nodes. 
We define size by allocating one unit to each lambda node and one to each application node.
This, for a given $N$,  will produce lambda terms of size $2N+1$.
But first we will only generate  term skeletons for which this constraint holds, with a dummy {\tt leaf} node. As these are a special case
of Motzkin trees (also called binary-unary trees,
see {\bf A001006} at \cite{intseq}),
we call them {\em linear Motzkin skeletons}.
We will obtain linear lambda terms by decorating these trees
with lambda binders and leaf variables in their scope.

\begin{code}
linear_motzkin(N,E):-succ(N,N1),linear_motzkin(E,N,0,N1,0).
\end{code}
\begin{code}
linear_motzkin(leaf,A,A,L,L).
linear_motzkin(l(E),A1,A2,L1,L3):-pred(L1,L2),linear_motzkin(E,A1,A2,L2,L3).
linear_motzkin(a(E,F),A1,A4,L1,L3):-pred(A1,A2),
  linear_motzkin(E,A2,A3,L1,L2),
  linear_motzkin(F,A3,A4,L2,L3).
\end{code}

Interestingly, they correspond to sequence {\bf A024489} in
\cite{intseq}, giving  {\tt 1, 6, 70, 1050, 18018, 336336 ...},
which has a closed formula  and originates from a geometric interpretation
similar in terms of constraints on graph nodes, but it is not noted as related
to lambda terms or their Motzkin skeletons.

\subsection{Decorating the Linear Skeletons}

Next we decorate the Motzkin skeletons with lambda nodes and variables, while ensuring that we generate only closed terms. We push the lambda binder to a stack from which
each variable will pick a binder having it in its scope. This  ensures that we generate closed lambda terms.

The predicate {\tt closed\_almost\_linear\_term} initializes the counter {\tt N} for
application nodes. They are propagated down to {\tt 0} 
with variables {\tt A1,..,An} through the recursive calls.
The counter  {\tt N1 = N+1}  for lambda binders is propagated with the variables {L1,..,Ln}.
The stack of variables {\tt Vs}, initially empty,
makes available the lambda binders to the leaf variables.
The stack grows when a lambda constructor {\tt l/2} is introduced.

\BX
Almost linear lambda tree, having the same number of lambda nodes
as leaves, but not paired two by two, 3 occurrences of X and one of Y being the exception).\\
\begin{center}
term: $( \lambda X.( \lambda Y.X ~X )~ \lambda Z.Z )~ \lambda U.U $ ~~~~~~ tree:
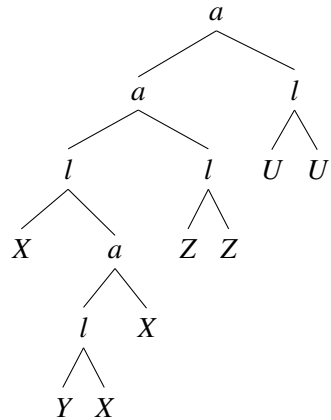
\Tree [.a [.a [.l [.X ] [.a [.l [.Y ] [.X ]  ] [.X ]  ]  ] [.l [.Z ] [.Z ]  ]  ] [.l [.U ] [.U ]  ]  ]
\end{center}
\EX

\begin{code}
closed_almost_linear_term(N,E):-succ(N,N1),
  closed_almost_linear_term(E,N,0,N1,0,[]).

closed_almost_linear_term(X,A,A,L,L,Vs):-member(X,Vs).
closed_almost_linear_term(l(X,E),A1,A2,L1,L3,Vs):-pred(L1,L2),
  closed_almost_linear_term(E,A1,A2,L2,L3,[X|Vs]).
closed_almost_linear_term(a(E,F),A1,A4,L1,L3,Vs):-pred(A1,A2),
  closed_almost_linear_term(E,A2,A3,L1,L2,Vs),
  closed_almost_linear_term(F,A3,A4,L2,L3,Vs).
\end{code}

Note that linearity constraints are only half-way enforced so far: 
we only ensure that the number of lambda nodes is equal to 
the number of variables they bind.

\subsection{Generating Closed Linear Lambda Terms}

To ensure that terms are linear, we will mark each lambda binder
when it reaches a variable. When exiting the expression
in the scope of the binder we test that it has been indeed marked. 
Note, that without this test we would obtain {\em affine} lambda terms.
The following predicates implement these operations.
\begin{code}
bind_once(V,X):-var(V),V=v(X).

check_binding(V,X):-nonvar(V),V=v(X).
\end{code}
Otherwise, the predicate {\tt linear\_lambda\_term} works
like {\tt closed\_almost\_linear\_term}.
\begin{code}
linear_lambda_term(N,E):-succ(N,N1),linear_lambda_term(E,N,0,N1,0,[]).

linear_lambda_term(X,A,A,L,L,Vs):-member(V,Vs),bind_once(V,X).
linear_lambda_term(l(X,E),A1,A2,L1,L3,Vs):-pred(L1,L2),
  linear_lambda_term(E,A1,A2,L2,L3,[V|Vs]),check_binding(V,X). 
linear_lambda_term(a(E,F),A1,A4,L1,L3,Vs):-pred(A1,A2),
  linear_lambda_term(E,A2,A3,L1,L2,Vs),
  linear_lambda_term(F,A3,A4,L2,L3,Vs).
\end{code}
This gives us the sequence {\bf A062980} in \cite{intseq}, starting as {\tt 1, 5, 60, 1105, 27120, 828250 ...}, confirming that they match results in \cite{lescanne18swiss,padl18}.

However, our goal is to generate unique theorems of a given size of linear implicational intuitionistic logic and that's on the other side of the Curry-Howard bridge. As otherwise the sizes of our lambda terms  can be smaller or larger than the formulas and more than one term can correspond to the same formula, we will need to restrict ourselves to {\em lambda terms in normal form}, i.e., terms not having lambdas on the left side of application nodes that could be simplified using $\beta$-reduction.


\subsection{Linear Normal Forms}

Generation of normal forms relies on  {\em neutral terms} that ensure that applications have as left nodes only variables or other application nodes. We ensure closedness and linearity constraints the same way as in the {\tt linear\_lambda\_term} generator.
\begin{code}
linear_normal_form(N,E):-succ(N,N1),linear_normal_form(E,N,0,N1,0,[]).

linear_normal_form(l(X,E),A1,A2,L1,L3,Vs):-pred(L1,L2),
  linear_normal_form(E,A1,A2,L2,L3,[V|Vs]),check_binding(V,X). 
linear_normal_form(E,A1,A2,L1,L3,Vs):-
  linear_neutral_term(E,A1,A2,L1,L3,Vs).

linear_neutral_term(X,A,A,L,L,Vs):-member(V,Vs),bind_once(V,X).
linear_neutral_term(a(E,F),A1,A4,L1,L3,Vs):-pred(A1,A2),
  linear_neutral_term(E,A2,A3,L1,L2,Vs),
  linear_normal_form(F,A3,A4,L2,L3,Vs).
\end{code}
Again, this gives us sequence {\bf A262301} in \cite{intseq} starting with
{\tt  1, 3, 26, 367, 7142, 176766, ...}, confirming the
results in \cite{lescanne18swiss,padl18}.
Again, generating the lambda terms in normal form is essential 
as otherwise multiple terms that $\beta$-reduce 
to the normal form would correspond to the same formula.

\subsection{Inferring the Types: Walking back Over the Curry-Howard Bridge }

Finally, we will also annotate our lambda terms with their inferred types. This is
quite easy as all linear terms are typable. Moreover, unlike in \cite{padl18},
unification does not require `occurs check'
as each lambda binds exactly one variable.

In fact, our type decoration mechanism can be seen as a simplified form
of the usual Hindley-Milner type inference \cite{hindleyTypes}
used to derive the types of
simply typed lambda terms, the main differences being that we 
do not need to use unification with occurs check and that 
we work exclusively on closed lambda terms.
\begin{code}
linear_typed_normal_form(N,E,T):-succ(N,N1),
  linear_typed_normal_form(E,T,N,0,N1,0,[]).

linear_typed_normal_form(l(X,E),(S '-o' T),A1,A2,L1,L3,Vs):-pred(L1,L2),
  linear_typed_normal_form(E,T,A1,A2,L2,L3,[V:S|Vs]),
  check_binding(V,X). 
linear_typed_normal_form(E,T,A1,A2,L1,L3,Vs):-
  linear_neutral_term(E,T,A1,A2,L1,L3,Vs).

linear_neutral_term(X,T,A,A,L,L,Vs):-member(V:TT,Vs),bind_once(V,X),T=TT.
linear_neutral_term(a(E,F),T,A1,A4,L1,L3,Vs):-pred(A1,A2),
  linear_neutral_term(E,(S '-o' T),A2,A3,L1,L2,Vs),
  linear_typed_normal_form(F,S,A3,A4,L2,L3,Vs).
\end{code}
The term counts, as expected, correspond to sequence {\bf A262301} in \cite{intseq},
under the title {\em number of normal linear lambda terms of size n with no free variables}.
Our Prolog program, runs the predicate {\tt linear\_typed\_normal\_form/3} 
in $O(N)$ space for terms of size $N$ and it generates billions of terms and types 
in a few hours, e.g., {\tt 1, 3, 26, 367, 7142, 176766, 5304356, 186954535, 7566084686}.
Note that in \cite{lescanne18swiss} a method to {\em count} linear terms analytically
provides counts up to higher values of $N$ for
their super-exponentially growing terms, but the corresponding
Haskell program\footnote{\href{https://raw.githubusercontent.com/PierreLescanne/CountingGeneratingAfffineLinearClosedLambdaterms/master/LinearNormalFormSize0or1.hs}{https://raw.githubusercontent.com/PierreLescanne/CountingGeneratingAfffineLinearClosedLambdaterms/master/Linear-NormalFormSize0or1.hs}}
 runs into memory problems after {\em }generating} 5304356 terms, even if given
250GB of memory.
The ability to go 3 orders of magnitude higher to  7566084686 actually generated terms (and even 4 if given
longer time or by parallelizing the generators as in \cite{tplp18}),
comes from the simple fact that Prolog recovers memory
on backtracking after each generated term is written out to a file or used
by another predicate.
Thus we work in $O(N)$ 
space for terms and formulas of size $N$, with no need for a garbage collector invocation.

\BX
Normal forms and their corresponding linear types.

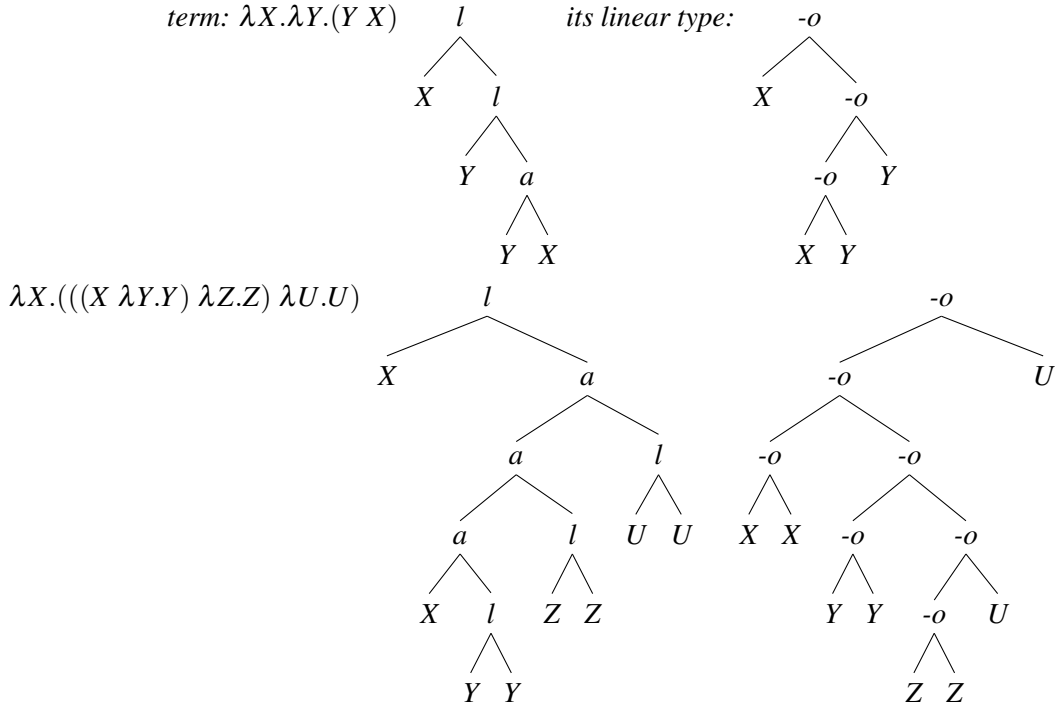
\begin{center}
term: $  \lambda X. \lambda Y.(Y ~X )$
\Tree [.l [.X ] [.l [.Y ] [.a [.Y ] [.X ]  ]  ]  ]
its linear type: \Tree [.-o [.X ] [.-o [.-o [.X ] [.Y ]  ] [.Y ]  ]  ]


$ \lambda X.(((X ~ \lambda Y.Y )~ \lambda Z.Z )~ \lambda U.U )$
\Tree [.l [.X ] [.a [.a [.a [.X ] [.l [.Y ] [.Y ]  ]  ] [.l [.Z ] [.Z ]  ]  ] [.l [.U ] [.U ]  ]  ]  ]~~~
\Tree [.-o [.-o [.-o [.X ] [.X ]  ] [.-o [.-o [.Y ] [.Y ]  ] [.-o [.-o [.Z ] [.Z ]  ] [.U ]  ]  ]  ] [.U ]  ]
\end{center}
\EX

\subsection{The Eureka Moment}
After looking at the generated terms and their types we observe the following surprising facts:

\BI
\I there are exactly two occurrences of each variable both in the theorems and their proof terms
\I theorems and their proof terms have the same size, counted as number of internal nodes
\EI 

{\em
Thus, we have solved the problem of generating  all tautologies size $N$ in the implicational fragment of propositional linear intuitionistic logic if the predicate {\tt linear\_typed\_normal\_form} implements a  generator of their proof-terms of size $N$,
for which the tautologies can be seen each as their principal type}.

It turns out that there's a {\em size-preserving 
bijection between linear lambda terms in normal form and their principal types}.
A proof of this follows immediately from \cite{zeilberger15} who attributes this observation to  \cite{mintsISO}. In \cite{zeilberger15} the bijection is proven by exhibiting a reversible transformation of oriented edges in the tree describing the linear lambda term in normal form, 
into corresponding oriented edges in the tree describing the linear implicational formula,
acting as its principal type.

{\em It follows that we have obtained a generator for all theorems of implicational linear intuitionistic propositional logic of a given size, as measured by the number of lollipops, without having to prove theorems, thus avoiding the need to call Turing-complete provers for linear logic or PSPACE-complete provers for propositional intuitionistic logic, simply by taking advantage of the existence of a {\em size-preserving bijection} between theorems and their corresponding proof terms and the Curry-Howard correspondence}.

Clearly, this is a ``Goldilocks'' situation, that, in a way, points out the very special case
that implicational formulas have in linear logic and equivalently, linear types have in type theory.
We plan future work on extending this result to the case of propositional affine linear logic, also known
to be decidable \cite{kopylov95}.

\subsection{Applications}

The  dataset containing generated theorems 
and their proof-terms in postfix form (as well as their LaTeX tree representations
marked as Prolog ``\%'' comments) that we make available at
\url{http://www.cse.unt.edu/~tarau/datasets/lltaut/}
can be used for correctness, performance and scalability testing
for linear logic theorem provers, in addition to the human made tests described in \cite{olarte2018},
as well as for providing similar tests for the Linear Haskell GHC 
compiler feature \cite{linhask}.

More importantly, the formula/proof-term pairs in the dataset
are likely to be usable to test if deep-learning systems
can perform a fairly interesting (and, in theory, learnable)
theorem proving task: if trained via a {\tt seq2seq} algorithm
on encodings of theorems and their proof-terms, can the resulting
model perform well on similar unseen formula/proof-term pairs? 
We have started   work in that
direction with promising initial results\footnote{
Our (successful!) experiments with training Recurrent Neural Networks using 
our implicational linear logic theorem dataset are available at:
 \url{https://github.com/ptarau/neuralgs} .
}.


\section{Discussion}\label{disc}

Proofs of implications in intuitionistic logic have long been recognized as fundamental, as they correspond to (closed) programs in functional programming. The same cannot be said about ``linear implications", as the tradition in linear logic tends to rewrite linear implications as multiplicative disjunctions (``pars" in proofnets) \cite{girard1987}. Proofnets, notwithstanding their logical appeal in simplifying proofs, are an ``acquired taste'', not shared by very many. One of our motivation for this research was using linear lambda-calculus \cite{gangoffour1993} to investigate both logical proofs in Intuitionistic Linear Logic and, further down the line, to investigate translations between it and intuitionistic logic proofs. 
Girard produced two such translations in his original paper on linear logic \cite{girard1987}. Applying these translations to well-known proofs in intuitionistic logic  (as for instance those described in the classic monograph \cite{kleene1952}) was a main motivation of \cite{olarte2018} leading to our initial interest for generating a benchmark of intuitionistic linear logic proofs, complementing the ones described in \cite{olarte2018}. 

But the hope for intuitionistic Linear Logic has always been to discover where duplication of hypotheses and, respectively, their erasure is safe, as far as the meaning of the proofs/programs is concerned. Our long term goal is to improve on the already known translations of intuitionistic logic into intuitionistic linear logic.  For that, we need to know more about the universe of existing linear proofs, like how many there are, their shapes, invariant properties, etc. Much work has already been done in this direction, see for example the work on "optimal reductions" \cite{gonthier1992} and on ``linear decorations"\cite{schellinx1994}. However, it seems fair to suggest that this work has not produced all the expected benefits, yet. The work described here is supposed to help with both of these aims.

\section{Related Work}\label{rel}

The classic reference for lambda calculus is \cite{bar84}.
The combinatorics and asymptotic behavior of various
classes of lambda terms are extensively studied in 
\cite{grygielGen}.
Distribution and density properties
of random lambda terms are described in \cite{ranlamb09}.
Asymptotic density properties of simple types (corresponding
to tautologies in implicational intuitionistic logic) have 
been studied in \cite{tautintclass} with the surprising
result that ``almost all'' classical tautologies are also 
intuitionistic ones.

The generation and counting of affine and linear lambda terms
is extensively covered in \cite{lescanne18swiss}, where, by using 
techniques from analytic combinatorics, 
much higher limits for {\em counting}  (but not {\em generating})
linear lambda terms are derived,
using efficient recurrence relations.
By contrast, our focus here is on generation. 
While producing also the corresponding tautologies,
our Prolog-based generators had actually go 3 orders of magnitude further than the Haskell program described
in \cite{lescanne18swiss}.

We have used extensively Prolog as a meta-language 
for the study of combinatorial and computational properties of lambda 
terms in papers like \cite{padl17,ppdp15tarau} covering different families
of terms and properties.
The idea of using types inferred for lambda terms as formulas 
for testing theorem provers  originates in \cite{padl19}.  
The current paper extends this line of research to linear logic, 
specifically to the  implicational fragment
of linear intuitionistic propositional logic.

The closest work  that we have used as starting point
for the intuitionistic logic prover is \cite{dy1}
describing the
{\bf LJT} calculus. 
Asymptotic behavior of linear and affine lambda terms, in
relation with the BCK and BCI combinator systems, as well as bijections
to combinatorial maps are studied in \cite{bodini13}. 
In \cite{lopst17temp}
analytic models are used to solve the problem of the asymptotic density of closable
and uniquely closable skeletons,
Motzkin trees that predetermine
existence and uniqueness of the closed lambda terms
decorating them.



The bijection between linear lambda terms in normal form and their
principal types, first proven in \cite{mintsISO} and explained also in terms of a geometric interpretation in \cite{zeilberger15}, has been instrumental in deriving the optimal final form
of our term/formula pair generator. 

\section{Conclusions}\label{concl}

We have derived declaratively  novel algorithms for the combinatorial generation of theorems in linear logic and their proof-terms.
The ability to declaratively encode constraints on the structure and the content of Prolog terms has enabled us to produce a generator for billions of theorems and their 
proof-terms in an important sublanguage of linear logic and to collect them into a a dataset usable for testing linear logic provers and training deep-learning systems 
for theorem proving, an emerging new task in machine learning.
By contrast to functional language implementations our algorithms fully recover space on backtracking, without even triggering Prolog's garbage collection.
This makes Prolog the language of choice for work exploring synergies between combinatorial generation, type inference and theorem proving.

\subsection*{Acknowledgments} 
We thank the anonymous reviewers of ICLP'2020 for their constructive comments and suggestions.


\bibliographystyle{eptcs}
\bibliography{tarau,theory,proglang,biblio}

\appendix

\section{The Implicational Intuitionistic Theorem Prover}\label{appendix}

\subsection{The LJT/G4ip Calculus 
}

Motivated by problems related to loop avoidance when implementing  Gentzen's {\bf LJ} calculus,
Roy Dyckhoff \cite{dy1}  introduces
the following rules for his {\tt LJT} calculus\footnote{Also called the G4ip calculus. Restricted here to the implicational fragment.}.\\\\
{\large
\noindent
\begin{math}
LJT_1:~~~~\frac{~}{A,\Gamma ~\vdash~ A}\\\\\\
LJT_2:~~~~\frac{A,\Gamma ~\vdash~ B}{\Gamma ~\vdash~ A\rightarrow B}\\\\\\
LJT_3:~~~~\frac{B,A,\Gamma ~\vdash~ G}{A \rightarrow B,A,\Gamma ~\vdash~ G}\\\\\\ 
LJT_4:~~~~\frac{D \rightarrow B,\Gamma ~\vdash~ C \rightarrow D ~~~~ B,\Gamma ~\vdash~ G}
{ \left( C \rightarrow D \right) \rightarrow B,\Gamma ~\vdash~ G }\\
\end{math}
}\\
The rules work with the context $\Gamma$
being either a multiset or a set.\\\\

\subsection{Extracting Proof Terms}

We refer to \cite{padl19} for the derivation steps leading from this calculus to Prolog-based theorem provers implementing it. We will focus here on extracting proof terms from a prover
adapted to cover the implicational fragment of propositional linear intuitionistic logic.

Extracting the {\em proof terms} (lambda terms having the formulas we prove as types) is achieved
by decorating the code  with application nodes {\tt a/2},
lambda nodes {\tt l/2} (with first argument a logic variable)
and leaf nodes (labeled with logic variables, same as the identically 
named ones  in the first argument of the corresponding {\tt l/2} nodes).

The fact that this is essentially the inverse of a type inference  algorithm 
(e.g., the Prolog-based one in \cite{hiking17})
points out how the decoration mechanism works.

\begin{code}
prove_ipc(T,ProofTerm):-prove_ipc(ProofTerm,T,[]).
prove_ipc(X,A,Vs):-memberchk(X:A,Vs),!. % leaf variable
prove_ipc(l(X,E),(A '-o' B),Vs):-!,prove_ipc(E,B,[X:A|Vs]).  % lambda term
prove_ipc(E,G,Vs1):- 
  member(_:V,Vs1),head_of(V,G),!, % fast fail if non-tautology
  select(S:(A '-o' B),Vs1,Vs2),   % source of application
  prove_ipc_imp(T,A,B,Vs2),!,       % target of application
  prove_ipc(E,G,[a(S,T):B|Vs2]).  % application
  
prove_ipc_imp(l(X,E),(C '-o' D),B,Vs):-!,prove_ipc(E,(C '-o' D),[X:(D '-o' B)|Vs]).
prove_ipc_imp(E,A,_,Vs):-memberchk(E:A,Vs). 

% optimization for quicker failure
head_of(_ '-o' B,G):-!,head_of(B,G).
head_of(G,G). 
\end{code}

Thus, lambda nodes decorate  {\em implication introductions} and
application nodes  decorate {\em modus ponens} reductions
in the corresponding calculus. Note that the two clauses of
{\tt  prove\_ipc\_imp} provide the target node $T$. When seen from
the type inference side, $T$ is the type resulting
from cancelling the source type $S$ and
the application type $S \rightarrow T$.

\begin{codeh}


sols_count(Goal, Times) :-
        Counter = counter(0),
        (   Goal,
            arg(1, Counter, N0),
            N is N0 + 1,
            nb_setarg(1, Counter, N),
            fail
        ;   arg(1, Counter, Times)
        ).

counts_for2(M,Generator,Ks):-
  findall(K,
  (between(0,M,L),
    sols_count(call(Generator,L,_),K),S is 2*L+1,ppp(size(L->S):count(K))),
  Ks).

counts_for3(M,Generator,Ks):-
  findall(K,
  (between(0,M,L),
    sols_count(call(Generator,L,_,_),K),S is 2*L+1,ppp(size(L->S):count(K))),
  Ks).
  
/*
?- time(counts_for(7,linear_typed_normal_form,Ks)).
Ks = [1, 3, 26, 367, 7142, 176766, 5304356, 186954535].
*/

ppp(X):-portray_clause(X).

ppt(X):-ppp(X).


show2(N,Gen):-call(Gen,N,X),ppt(X),nl,fail;true.
show3(N,Gen):-call(Gen,N,X,T),ppt(X),ppt(T),ppp('------'),nl,fail;true.


gt(N):-counts_for3(N,gen_taut,Ks),ppp('LinearTautologies'(N)=Ks).

mc(N):-counts_for2(N,linear_motzkin,Ks),ppp('LinearMotzkin'(N)=Ks).

lc(N):-counts_for2(N,linear_lambda_term,Ks),ppp('LinearLambda'(N)=Ks).

lt(N):-counts_for3(N,linear_typed_normal_form,Ks),ppp('LinearTermsAndType'(N)=Ks).

tr(N):-counts_for3(N,test_reversible_prover,Ks),ppp('LinearTermsAndType'(N)=Ks).
go:-gt(7),nl,mc(5),nl,lc(5),nl,lt(5),nl,tr(3).
\end{codeh}

\end{document}